\documentclass[lettersize,journal]{IEEEtran}
\usepackage{amsmath,amsfonts}
\usepackage{algorithmic}
\usepackage{algorithm}
\usepackage{array}
\usepackage[caption=false,font=normalsize,labelfont=sf,textfont=sf]{subfig}
\usepackage{textcomp}
\usepackage{stfloats}
\usepackage{url}
\usepackage{verbatim}
\usepackage{graphicx}
\usepackage{cite}
\usepackage[numbers,sort,compress]{natbib}
\bstctlcite{IEEEexample:BSTcontrol}
\begin{document}

\title{PDA-LSTM:~Knowledge-Driven LSTM Data Arrangement Optimization for 3D NAND Flash Storage with Lateral Charge Migration Suppression}

\author{Qianhui~Li,
 Weiya~Wang, 
 Qianqi~Zhao,
 Tong~Qu,
 Jing~He, 
 Xuhong~Qiang$^{a,b}$,\\
 Jingwen~Hou,
 Ke~Chen,
 Bao~Zhang$^{*}$, 
 Qi~Wang$^{*}$
\thanks{Qianhui Li, Weiya Wang, Jingwen Hou are with the Institute of Microelectronics of the Chinese Academy of Sciences, Beijing 100029, China (e-mail:liqianhui@ime.ac.cn).  Qianqi Zhao, Tong Qu, Jing He, Ke Chen, Bao Zhang, Qi Wang, Tianchun Ye are with the Institute of Microelectronics of the Chinese Academy of Sciences, Beijing 100029, China, and the University of Chinese Academy of Sciences, Beijing 100049, China.}

}

\maketitle

\begin{abstract}
LSTM, a neural-symbolic framework integrating Long Short-Term Memory networks with combinatorial optimization, is a traditional model squence tasks. Squence task includes data arrangement for storaging improvement in 3D NAND FLASH. The target arrangement should reduce critical challenges from lateral charge migration (LCM)-induced retention errors for multi-bit cell storage, particularly in high-density QLC (4-bit/cell) designs. While conventional rule-based data mapping methods (e.g., WBVM, DVDS) focus on intra-page optimization, their heuristic strategies fail to address dynamic inter-page data dependencies. Thus, we proposed PDA-LSTM to arrange intra-page data for LCM suppression. PDA-LSTM applies a long-short term memory (LSTM) neural network to compute a data arrangement probability matrix from input page data pattern with the loss function based on relationship of storaged data arrangement and LCM effect. The arrangement is to minimize the global impacts derived from the LCM among wordlines. Since each page data can be arranged only once, we design a transformation from output matrix of LSTM network to nonrepetitive sequence generation probability matrix to assist training process. Finally, we acquire an optimal data mapping table according to the output matrix of LSTM to implement data redistribution.  PDA-LSTM demonstrates AI-centric advantages by eliminating manual flag-bit designs (12.7\% metadata reduction) and exhibiting 89.4\% cross-architecture (QLC) policy validity. Experiments on 512Gb QLC chips show 80.4\% average BER reduction versus non-optimized baselines, outperforming WBVM and DVDS by 18.4\% and 15.2\% respectively under 64-byte codes, with ablation studies confirming LSTM-based correlation modeling contributes 63.8\% of BER improvement. 
\end{abstract}


\section{INTRODUCTION}
\IEEEPARstart{L}{STM} is a common neural network for sequence related tasks and it can be used in computer-aided design. With the development of the big data era, processing of massive data in the storage system device has become an important need for large-scale computer systems\cite{kim2021cmos}. 
Solid-state drivers (SSDs) serve as peripherals assisting hosts in large-scale data management. As the data storage medium of SSDs, 3D NAND flash encounters reliability issues such as endurance degradation, cell-to-cell interference, and data retention errors, leading to an increase in read errors\cite{8013174}. Quarter level cell (QLC) 3D NAND flash increases storage density by expanding the threshold voltage intervals, leading to narrower read windows and increased read errors\cite{9006406}\cite{liu_qlc_2017}\cite{8739689}. Consequently, this affects the I/O read/write operations between SSDs and hosts, significantly impacting the performance of QLC based large-scale data storage systems\cite{9631819}.
\begin{figure}
    \centering    
    \includegraphics[width=1\linewidth]{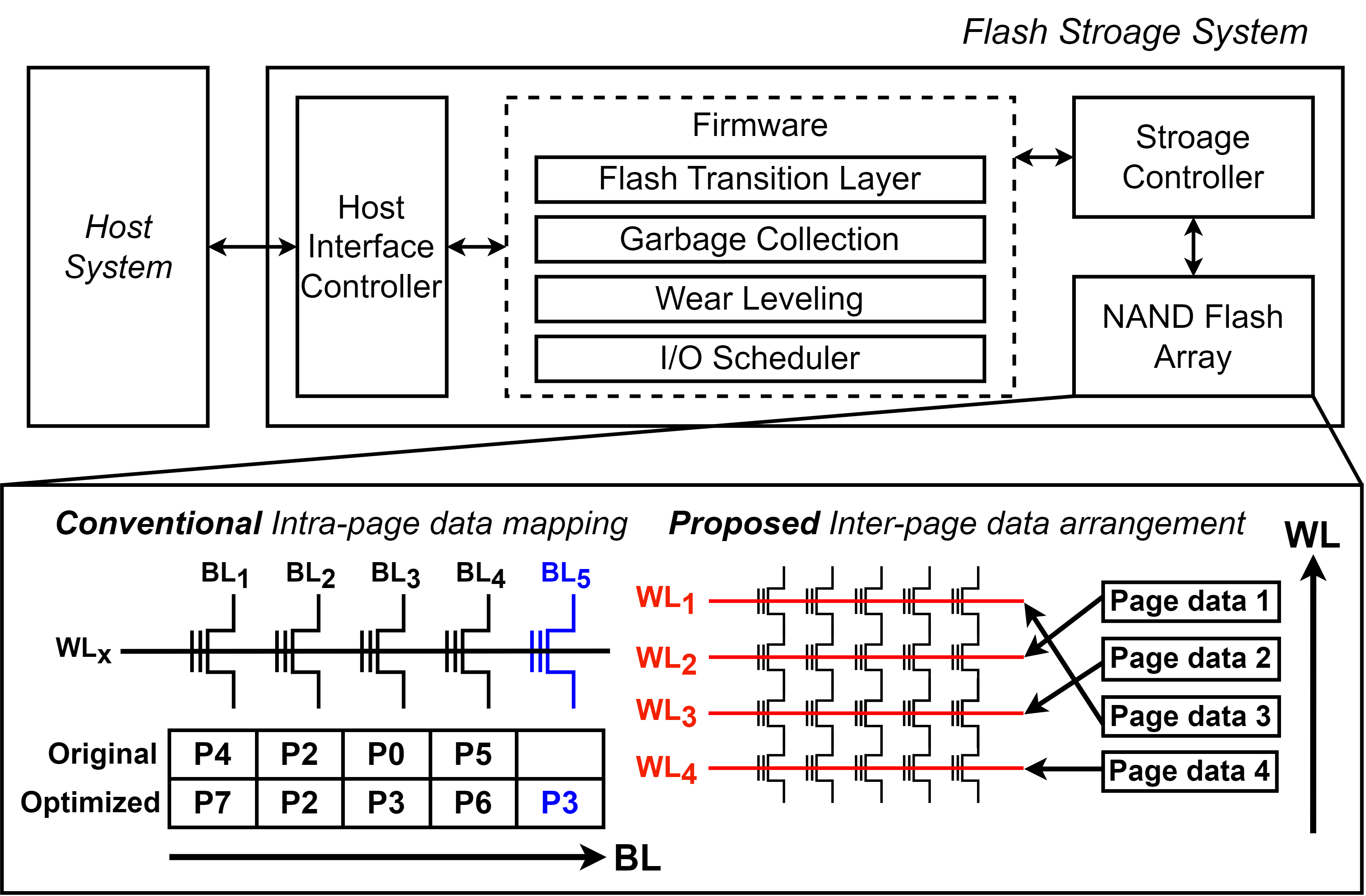}

    \caption{Brief view of flash-based storage system and concept of this paper}
    \label{fig:enter-label}
\end{figure}
\par
To improve the performance of SSD system, previous works have dedicated significant efforts in read error reduction. 
For the source of the retention error, Lateral charge migration (LCM)\cite{maconi2012comprehensive}\cite{wang2020lateral}, unique to NAND flash memory with 3D stacking structure, contributes to read errors significantly for QLC 3D NAND flash memory. Previous research has proposed methods such as WBVM\cite{deguchi2017word}\cite{deguchi2018write} and DVDS\cite{suzuki2018endurance} to add extra flag bits for intra-page data mapping to alleviate LCM effect\cite{marx_big_2013}\cite{6547630}\cite{7123563}. 
However, intra-page data mapping methods require a large number of redundant flag bits to mark the mapping information of threshold voltage distribution states, and without considering from the perspective of the position arrangement of page data. 
\par
Based on the observation of page data sorting for LCM suppression, we propose the page data arrangement (PDA) method based on long-short term memory (LSTM) neural network to reduce LCM effect. The proposed PDA-LSTM learns the neighbor relationship of threshold voltage states and achieves the intra-page data arrangement. PDA-LSTM computes a data arrangement probability matrix from input page data to minimize the global impacts derived from the LCM among wordlines. Since each page data can be arranged only once, a transformation from output matrix of LSTM network to nonrepetitive sequence generation probability matrix is designed to assist network training process. Training process is with the loss function based on the physical mechanism about LCM effect evaluation function for different storaged data arrangement.
To characterize the LCM effect of input data, a triple combination score matrix based on page data patterns is designed in network training processes. LCM effect Score calculation of input page data is a time-consuming operation, primarily due to the large input data volume. Fortunately, LCM evaluation score calculation only occurs in the training process, which is not required during the inference process. In the inference process, an optimal data address mapping table according to the output matrix of PDA-LSTM is used to alleviate LCM effect to decrease retention errors. 
\par
The contributions and exploit can be summarized as follows. 
\begin{itemize}
    \item Proposed the arrangement strategy from page data to storage position for the first time, to our knowledge. 
    \item Established the scoring rules for triple adjacent wordlines combination in QLC based on LCM appearance.
    \item Use LSTM network to embed the threshold voltage states distribution of the input data and provide the best mapping of page data and storage positions through linear and softmax layer. The mapping makes LCM function approach the minium.
    \item After completing the network training, LCM evaluation score calculation, which is a time-consuming operation, is not required in the inference process. 
    \item The method can not only decrease the LCM effect to random data, but also nearly pertains to any reality data, such as audio, image, video, text, etc.
    \item The method only needs  optimal data mapping table can be implemented by address mapping in FTL according to the output of PDA-LSTM to achieve page data arrangement and LCM alleviation. No redundant 3D NAND flash storage space needs.
\end{itemize}
\par
The rest of this paper is organized as follows. Section II introduces background and motivation. The proposed LCM suppression method and detailed designs
of PDA-LSTM is presented in Section III and IV. Section V presents the experimental
results and Section VI concludes this paper.

\section{BACKGROUND AND MOTIVATION}
\subsection{3D NAND flash memory architecture}
Silicon-Oxide-Nitride-Oxide-Silicon (SONOS) \cite{maconi2012comprehensive,ouyang2020optimization} is one of the most widely used 3D NAND flash storage unit structures due to its excellent data retention performance. SONOS achieves data storage by programming methods. These methods control the quantity of trapped charges in the nitride layer, thereby altering the threshold voltage of 3D NAND flash storage units. 
A pivotal advancement in augmenting the data storage capacity of 3D NAND flash memory is the adoption of multi-bit cell technology. This approach discretizes the threshold voltage range into $m$ distinct intervals, enabling the representation of $n$ bits per cell, where $m$ is defined by the equation $m = 2^n$. Here, $n$ represents a positive integer. For example, Quad level cell (QLC) \cite{khakifirooz202130} achieves 4-bit storage ($n=4$) by dividing the threshold voltage into 16 ($m=16$) intervals. 
\par
\begin{figure*}
    \centering  
    \includegraphics[width=0.8\linewidth]{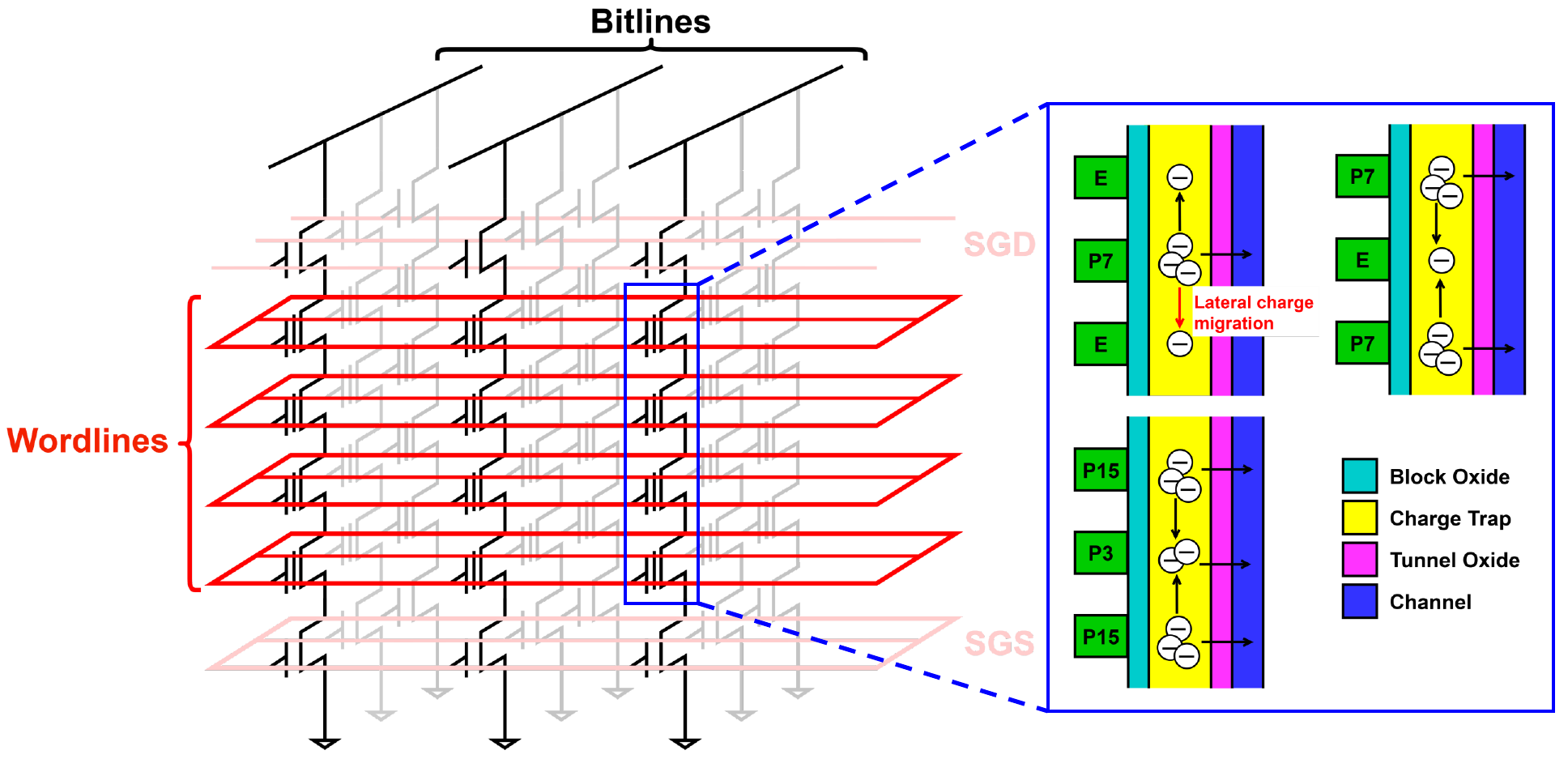}
    \caption{QLC 3D NAND flash and lateral charge migration}
    \label{fig:NAND_structure}
\end{figure*}

The vertical stacking of storage units is another key technology for achieving high-density storage in 3D NAND flash. The array structure of a block within 3D NAND flash is depicted in Fig. \ref{fig:NAND_structure}. The block serves as both the smallest array structure unit and the minimum unit for erasure in 3D NAND flash. By stacking wordlines (WLs), 3D NAND flash achieves its remarkable three-dimensional structure and fabricates storage units within channel holes that traverse all stacked word lines. A string is formed by all storage units located within the same channel hole. The storage units in different strings are connected to bit lines (BLs) through select gates, typically with 4-6 strings connected to BLs within a block. Within a word line, storage units located on different BLs collectively constitute a physical page, which is typically sized at 18KB. Additionally, within a block, the number of strings connected to a BL matches the number of physical pages on a WL. For the storage capacity of a single storage unit, denoted as n, a physical page can be further divided into n logical pages. 
\subsection{Lateral charge migration}
In 3D NAND flash memory, LCM \cite{peng2020impacts} refers to the phenomenon where electric charge stored in memory cells drifts laterally within the memory cell array. This lateral movement of charge can occur over time due to various factors such as temperature fluctuations, electrical stress, and material defects\cite{mizoguchi2017lateral}. As charge migrates laterally within the memory cell array, it can interfere with neighboring cells, leading to unintended changes in stored data\cite{6044201}. LCM effect induces a shift in the threshold voltage of flash memory, consequently leading to an increase in read errors\cite{Wang_2020}. For QLC 3D NAND flash memory, the increased number of voltage levels makes storage cells more susceptible to LCM compared to cells with fewer states\cite{8720566}. Therefore, QLC 3D NAND flash is more susceptible to the LCM effects, resulting in a higher occurrence of read errors. In addition, LCM can be influenced by the data pattern stored in the NAND flash memory\cite{8353632}. Therefore, LCM can be mitigated by employing data mapping or arrangement techniques, thereby reducing data read errors \cite{deguchi2018write}\cite{suzuki2018endurance}. 

\subsection{Intelligent sequence generation algorithms}
To get the optimal data arrangement, many ways are considered. However, without intelligent algorithms, it should spend a huge times to get the proper arrangement, So find a intelligent method is necessary to get the arrangement rapidly. The intra-data arrangement can be regarded as the address sequence generation task. The referenced intelligence methods includes TSP, Greedy and LSTM network.
\par
\textbf{Travel Salesman Problem (TSP) solutions.} 
TSP is a classical combinatorial optimisation problem. The sequence generation problem can be viewed as a kind of TSP because the optimization objective is a particular metric. The classic TSP can be described as follows: a salesman has to go to several cities to sell his goods, and the salesman starts from one city and needs to return to the starting place after passing through all the cities\cite{junger1995traveling}. Provide a travel route to make the total journey shortest. The traditional methods of solving the TSP are mainly genetic algorithms, simulated annealing, ant colony algorithms, forbidden search algorithms, and greedy algorithms. Among them, simulated annealing (SA) is to accept a worse solution than the current solution with a certain probability, and then use this worse solution to continue searching for the advantage in the local optimal solution\cite{malek1989serial}. SA is widely used in TSP As for its ability to probabilistically jump out and eventually converge to the global optimum. Greedy strategy is the nearest-neighbour, starting from any city, each time in the city that has not been to the city to choose the closest to one of the cities that have been selected, until it has passed through all the cities. It is also a commonly alternative solution because it can give an approximate optimal solution with less computational complexity\cite{karabulut2014variable}. 
\par
\begin{figure}
    \centering  
    \includegraphics[width=1\linewidth]{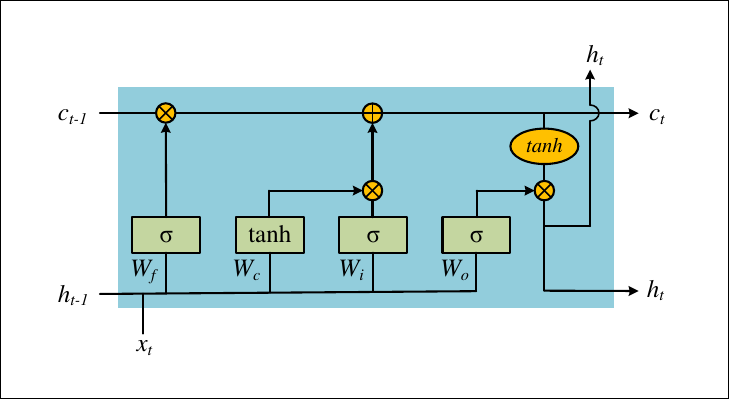}
    \caption{Long-short term memory unit}
    \label{fig:lstm unit}
\end{figure}

\textbf{Long-short term memory (LSTM) network.} 
In some sequence generation tasks, LSTM model was invented as a specialized variant of RNN, which overcomes difficulties when handling long sequences with RNN network\cite{graves2012long}. So it can capture long-term sequential dependencies more effective. This network introduces memory cells that allow information to be stored and retrieved over extended periods. Additionally, LSTM incorporates gating mechanisms, including the input gate, forget gate, and output gate, to regulate the flow of information within the LSTM unit as shown in Fig. \ref{fig:lstm unit}. The input gate controls the influx of new information into the memory cell, the forget gate determines which information to discard, and the output gate determines which information to output\cite{hochreiter1997long}. By incorporating memory cells and gating mechanisms, LSTM effectively captures long-term dependencies in sequential data, making it suitable for tasks involving varying sequence lengths\cite{van2020review}. 

\subsection{Motivation   }

LCM effect presents significant challenges in storage systems. Traditional methods for mitigating this effect often require a large number of identification bits, resulting in excessive storage resource consumption. Furthermore, these methods fail to consider the impact of data arrangement within the same string on LCM. To address these limitations, this paper proposes the use of LSTM network to optimize data arrangement and alleviate the LCM effect in 3D NAND flash.  This approach enables a more comprehensive understanding of the relationship between data arrangement patterns and the LCM effect, facilitating the identification of optimal arrangements. 

\section{PAGE DATA ARRANGEMENT TO ALLEVIATE LCM}
Based on the relation of charge migration and data patterns, we propose a data arrangement strategy to alleviate LCM to reduce retention errors\cite{9241228}. In this strategy, reflection of storage location and page data is determined by the optimal arrangement with minimal LCM effect and the LCM effect is quantified into scores.
\subsection{Quantitative representation of LCM}
LCM is affected by the threshold voltage difference between adjacent 3D NAND flash cells on the same string. The larger the threshold voltage difference between adjacent storage cells on the same string, the stronger the lateral charge migration and the larger the threshold voltage shift and BERs \cite{8936476}\cite{6392183}\cite{9241228}. Threshold voltage difference for $x_n$ on upside and underside cells is calculated with $|x_n-x_{n+1}|$ and $|x_n-x_{n-1}|$ respectively. Where $x_n$ ranges from 0 to 15, representing threshold voltage distribution levels. 
\par
Even if the threshold voltage difference is the same, LCM effect of the intermediate cell by the upside and underside cells is different. Experiments demonstrated that the intermediate cell is more affected by the underside cell, which is about four times as much as that of the lower cell \cite{8310323}\cite{8895844}. The influence coefficients of upside and underside are $k_1$ and $k_2$, respectively. The ratio of $k_1$ to $k_2$ is $4$ to $1$ \cite{mizoguchi2017lateral}. 
\par
Affected by lateral electric field, the saturation value of the threshold voltage offset  caused by LCM becomes larger as the program level increases \cite{pasuto2013lateral}. Therefore, we use the level of threshold voltage $x_n$ representing the saturation coefficient. Erased states typically residing within the negative voltage range exert a greater influence on LCM effects. Therefore, we specifically define $Ae$ as the coupling impact coefficient of the adjacent erased states.
\par

\subsection{LCM effect based data score}
Based on the principle that a higher score corresponds to a smaller LCM effect and lower error rates, we establish scoring rules that delineate the impact of neighboring cells on the intermediate cell. The score of the intermediate cell can be calculated as
\begin{align}
     s(x_{n-1},x_{n},x_{n+1})=A_e  (16-x_{n}) \cdot f(x_{n-1},x_{n},x_{n+1}) \\
     f=\frac{k_2(16-|x_{n-1}-x_{n}|)+k_1(16-|x_{n}-x_{n+1}|)}{\alpha (k1+k2)}.\\
     \label{equation:score}
\end{align}

Where $(16-x_{n})$, $(16-|x_{n-1}-x_{n}|)$, $(16-|x_{n}-x_{n+1}|)$ represent formula terms associated with saturation coefficient, underside cell effect, and upside cell effect, respectively \cite{suzuki2018endurance}. The coefficient $\alpha$ is used to control the range of scores. Table \ref{tab:value_Ae} displays the values of $Ae$. $Ae$ means the coupling interaction between the upper and lower cells and the central cell. If $x_n$ is in the E state, then the LCM influence of $x_{n-1}$ and $x_{n+1}$ on $x_n$ is not significant, resulting in a higher score. If $x_n$ is in the P state and the two adjacent cells are in the E state, then the LCM influence is greater, resulting in a lower score.  Where 0 represents the erase state and $P_{x_{n-1}}$ represents the underside cell is in the program state with level $x_{n-1}$.
\begin{table}[h]
    \centering
    \caption{$A_e$ for triple adjacent threshold voltage state combination}
    \begin{tabular}{c c c|c}
    \hline
       $x_{n-1}$ & $x_{n}$ & $x_{n+1}$  & $A_e$ \\
    \hline
    $P_{x_{n-1}}$ & 0 &$P_{x_{n+1}}$ &      5  \\
     0  &  0  &  0 &      5  \\
     $P_{x_{n-1}}$ &$P_{x_{n}}$ &$P_{x_{n+1}}$ &      5\\
     0  & $P_{x_{n}}$  & 0  &      1  \\
     0 & 0 &$P_{x_{n+1}}$ &      5  \\
     0 &$P_{x_{n}}$& $P_{x_{n+1}}$ &      2  \\
     $P_{x_{n-1}}$ & 0 & 0 &   5  \\
     $P_{x_{n-1}}$ & $P_{x_{n}}$ & 0 &      2  \\
    \hline
    \end{tabular}
    \label{tab:value_Ae}
\end{table}
\par
Based on the scores of adjacent cells located on the same string in equation (\ref{equation:score}), all cell scores in different bitlines are summed to obtain the page data score. Then, the page data scores of the adjacent wordlines affected by LCM effects are also summed to obtain the LCM evaluation score for data pattern, which is calculated by
\begin{equation}
S_{T}=\sum_{j \in WL_n} \sum_{i \in  BL_n} s_{(i,j)}(x_{n-1},x_{n},x_{n+1}).
\label{equation:score_total}
\end{equation}
\par
\begin{figure}
    \centering
    \includegraphics[width=0.5\linewidth]{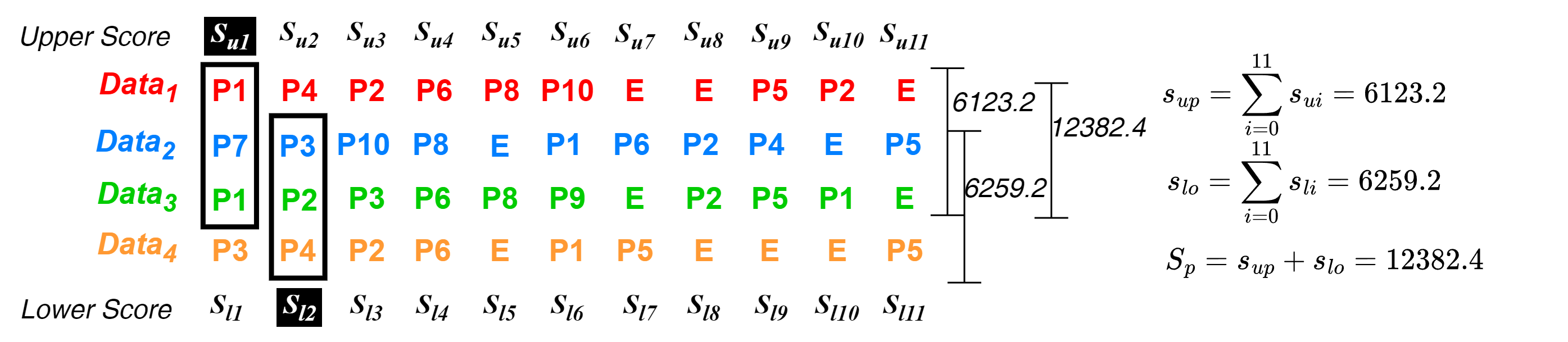}
    \caption{Arrangement example for data in four adjacent physics pages sharing the same bit lines.}
    \label{fig:LCM_example}
\end{figure}

The higher the LCM evaluation score, the lower the LCM effect, and the lower the data retention errors. As shown in Fig. \ref{fig:LCM_example}, the proposed page data arrangement strategy sets the page data in a reasonable physics position to maximize the total scores to alleviate LCM. After the arrangement process, the rank of page data may difference from the previous, so it can also be regarded as a data reordering process. In the upcoming sections, we will delve into the methodology of page data arrangement based on the LSTM neural networks. 
\begin{figure}
    \centering
    \includegraphics[width=0.5\linewidth]{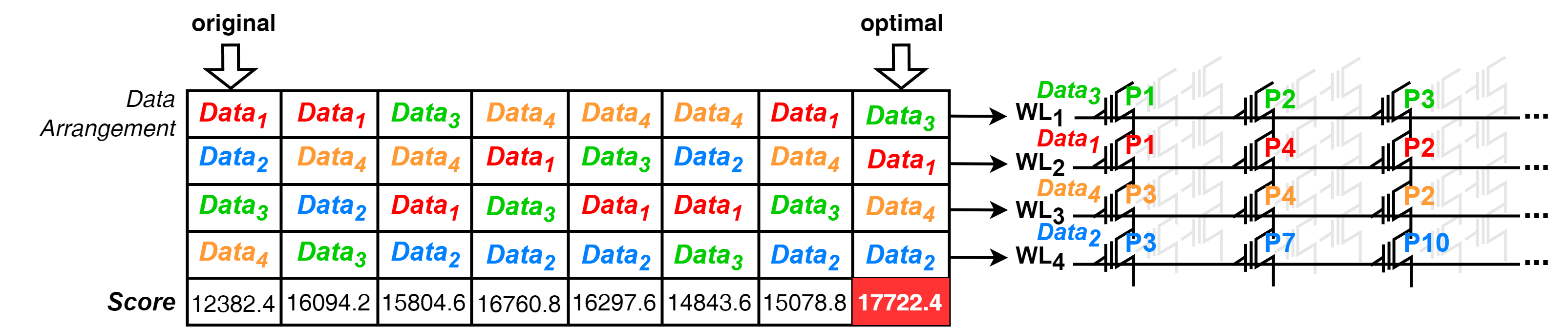}
    \caption{Assuming that there are four page data Data1, Data2, Data3, Data4 to be written, there are 24 placement strategies for these four data. Taking the data programming status of each bitline randomly selected as an example, (a) gives the comprehensive score of the cells on all bitlines in each placement method according to the LSM evaluation strategy, and selects the best data placement method as shown in (b).}
    \label{fig:sub2}
\end{figure}

\section{LSTM-BASED PAGE DATA ARRANGEMENT}
Page data arrangement strategy we proposed to alleviate LCM impact can be converted as the sequence generation task without repeating. Based on LSTM network widely use in sequence generation tasks, we proposed PDA-LSTM network to achieve page data arrangement. 
\subsection{LSTM network design}
\begin{figure*}
    \centering    \includegraphics[width=0.95\linewidth]{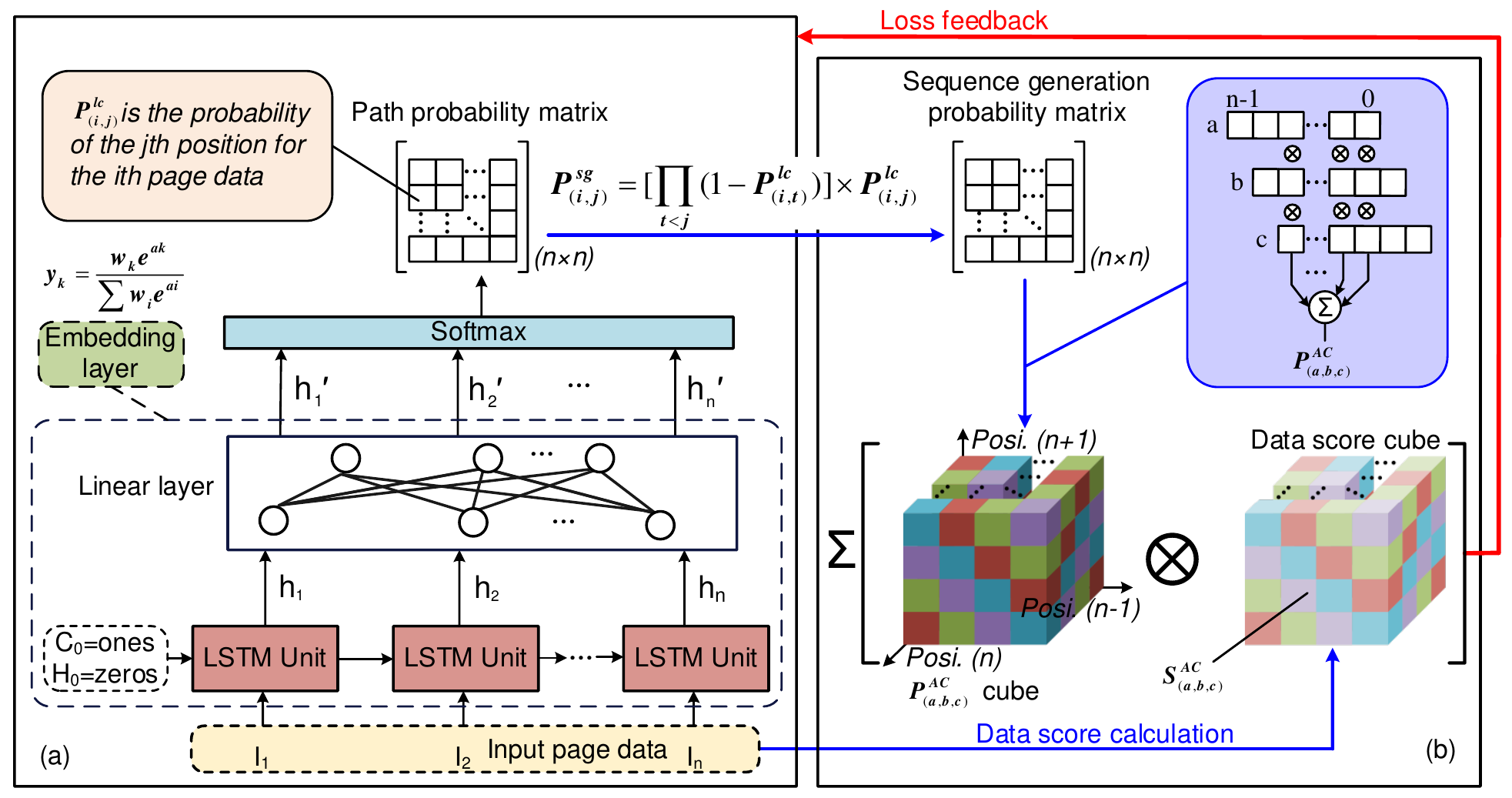}    \caption{\textbf{PDA-LSTM training network.}  The model network is shown in (a), consisting of input data pattern, LSTM embedding network. In model training process in (b), location probability distribution matrix, as the output of PDA-LSTM network, is converted to non-repeated sequence generation probability matrix and page triple combination probability calculation. The loss function is obtained by multiplying the score of each combination by the sum of its actual combination probability.}
    \label{fig:total PDA-lstm}
\end{figure*}
The proposed PDA-LSTM network is consisted of three parts, embedding layer with LSTM, linear layer with full connection, and classify layer with softmax network as displayed in Fig. \ref{fig:total PDA-lstm} (a). 
\par
\textbf{Embedding layer with LSTM.}
In QLC NAND flash, four logical page data consists a physical page data. 
Based on the Gray code of QLC NAND flash, the physical data elements can be depicted as the integer with domain in 0 to 15. The input of PDA-LSTM is all QLC physical page data, whose elements limited in integer domain of 0 to 15 to depict all QLC programming states. So the input length of the model is the number of physical pages, the dimension is the number of cells in each physical page. In embedding input data process, we apply LSTM network which the next input is relative to the former output. This character ensures its advantage in the page data arrangement task. 
\par
In LSTM processing of normal data input, the data are usually first deal with layer normalization\cite{9871606}\cite{8765616}.
However, the layer normalization of QLC physical page data may disrupt the distribution states relation among different page data. Therefore, we choose not to do the layer normalization to the input data. 
\par
\textbf{Linear layer.}
The Linear layer is used to process the embedded feature of the hidden layer. It is a fully connection neural network and connects the LSTM embedding layer for input and the softmax layer towards output\cite{9758404}. So it can determine the output result in some extend and its layer number can both influence the calculation cost and effect of the whole network.
\par
\textbf{Softmax layer.}
The softmax layer is a common network for classify tasks. The output of this layer is the classify probability distribution of each sample. The algorithm for softmax is stated in equation (\ref{equation:softmax}).
\begin{equation}
y_k=\frac{\boldsymbol{W_k} \times e^{a_k}}{\sum{\boldsymbol{W_i} \times e^{a_i}}}
\label{equation:softmax}
\end{equation}
\subsection{Loss calculation based on semi-supervisor learning}
To train the proposed model achieving optimal page data arrangement, since data-driven needs the best arrangement of different data, which should be difficult to get. So loss function design play a key role in optimizing the model towards the optimal sequence generation. Designing the loss function induces the semi-supervisor learning concept.

Semi-supervised learning is a machine learning technique that uses a combination of large amounts of unlabeled data (often several times the size of labeled data) with a small amount of labeled data for model training. In this scenario, the network training should be driven by physics knowledge, which represented in loss function.

The optimizing proposal can be divided into two sections, one is factories in the generated sequence non-repeated, the other is ensuring the total combination score to reach max. So when designing the loss function, we introduce probability related matrix $\boldsymbol{P^{sg}}$,$\boldsymbol{P^{AC}}$ and score related matrix $\boldsymbol{S^{AC}}$ to assist model training as shown in Fig \ref{fig:total PDA-lstm} (b).

\par
\textbf{Sequence generation probability design to assist non-repeated arrangement.} The output matrix $\boldsymbol{P^{lc}}$ of the PDA-LSTM network describes position probability for each physical page data. $\boldsymbol{P^{lc}}$ represents position probability, which is an input for loss calculation.Restricted to each physical page data can only occupy one position, the final position probability matrix should meet non-repeated generation rule. As a result, we design the sequence generation probability matrix in training process according to conditional probability related to the former generated elements. It is to say, if the probability for former (n-1) positions generating page j is $p_{0j}$,$p_{1j}$....$p_{(n-2)j}$ and the the prior probability for n$\bold{th}$ position generating page j is the product of $1-p_{0j}$,$1-p_{1j}$....$1-p_{(n-2)j}$, as it shown in equation \ref{equation:sg}. 
Based on the prior for rank of each page, the sequence probability matrix should be the multiply of the prior and the position probability matrix.
\begin{equation}    P_{(i,j)}^{sg}=\left\{
\begin{aligned}
&\prod \limits_{t=0}^{j}(1-P_{(t,j)}^{path}) \cdot P_{(i,j)}^{path}, n>0\\
&P_{(i,j)}^{path}  ,  ~~~~~~~~~~~ n=0
\end{aligned}\right.\label{equation:sg}
\end{equation}
\par
\textbf{Combination probability matrix design.}
According to equation \ref{equation:score_total}, the 
evaluation score is closely correspond to each adjacent triple physical page combination. So we should calculate the probability of each adjacent triple combination calculation and get a combination probability matrix. Due to each word line in a combination is independent and each triple combination can be set in (N-2) potential positions when there are N word lines. The element of combination probability matrix is calculated by
\begin{equation}    
P_{a,b,c}^{AC}=\sum_{t=0}^{N-2}P_{(t,a)}^{sg}\times P_{(t,b)}^{sg}\times P_{(t,c)}^{sg}.
\label{equation:ac}
\end{equation}
\par
\begin{figure}[htbp]
    \centering    \includegraphics[width=1.0\linewidth]{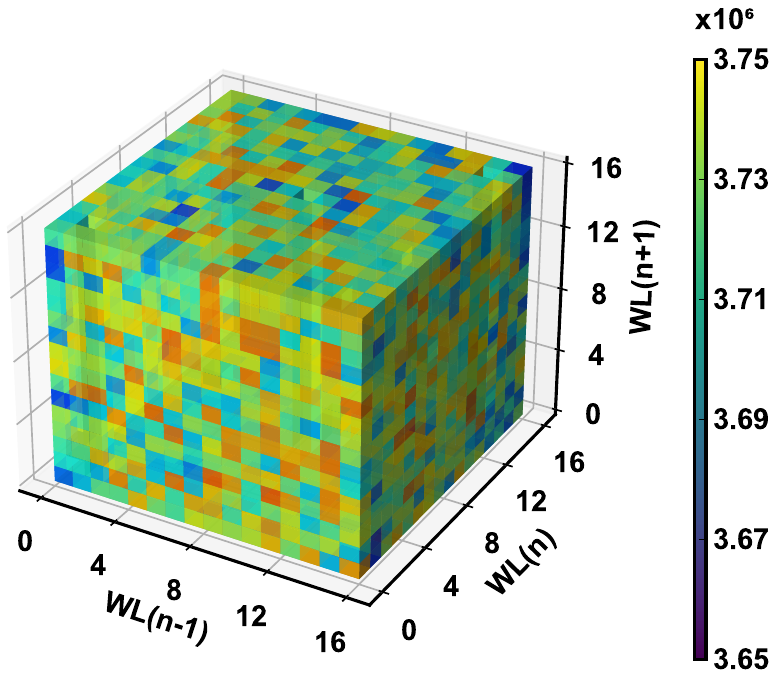}
    \caption{Adjacent combination score matrix}
    \label{fig:sac cube}
\end{figure}

\textbf{Adjacent combination score matrix design.}
The score matrix is design foundation to the lateral charge loss evaluation function, which is related three physical pages in adjacent wordlines. So it puts three adjacent physical pages as a combination group. And each combination reflection a score. If there are N word lines, there should be $N \times N \times N$ triple combinations and according to non-repeated regulation, wordlines in a combination can not be the same. So only $N\times(N-1)\times(N-2)$ triple combinations is valid. And invalid combinations was set the score to 0 while that in valid combinations was set as $s(x_a,x_b,x_c)$. 
\begin{equation}  
S^{AC}_{(a,b,c)} = \left\{
\begin{aligned}
& s(x_a,x_b,x_c),a\neq b \neq c\\
& 0,~~~~~~~~~~~~~~~~others
\end{aligned}\right.\label{equation:Sac}
\end{equation}
As a training target, the network should learning the relationship between score and data pattern feature. So we design a triple adjacent combination score matrix $\boldsymbol{S^{AC}}$, each element stands for each triple pages data combination LCM evaluation score as shown in Fig. \ref{fig:sac cube}. This matrix can be regarded as the label in the training process. 
\par
\textbf{Sum-score based on probability}
With adjacent combination probability matrix $\boldsymbol{P^{AC}}$ and score matrix for each triple adjacent combination $\boldsymbol{S^{AC}}$. Since each triple adjacent combination has (N-2) potential placement. The total score should be the sum of all of the (N-2) adjacent triple combination. Foundation on the probability,the final score can be described as equation \ref{Ssum}.
\begin{equation}    
S_{m}=\sum_{t=0}^{N-2}P_{(i,j,k)}^{ac}\times S_{(i,j,k)}^{ac}
\label{Ssum}
\end{equation}
\par
The higher score, the better pages in various wordlines sorting. However, Considering the optimization for model with loss, it is that the model optimization is always with the loss reduction. Therefore, we set the loss function as the negative value of final score. 
\begin{equation}    
Loss=-S_{m}
\end{equation}

\subsection{Train and test}
The model training can produce the proposed PDA-LSTM network, which can provide the best rank for physical page data by inference. The train and test set are generated random data.
\par
\textbf{Dataset production.} 
Since we have analysed representation for random data. So the dataset is random data limited in 0 to 15 with the size $16\times(8\times 1024\times 18)$. Here we set 16 pages in respective wordlines. Capacity of each page is 18KB, so the cells in each page should be $18\times1024\times8$. The ratio of the number of training sets to the number of test sets is 7:3. 
\par
\textbf{Scoring learning in training.}
During the training process, the model should learn which kind of arrangement for input data can bring the highest total score and non-repeated rule.


\textbf{Featuring learning target.}
The generation should be non-repeated,which means the output probability for each rank destination page should as close to 1 as possible while others as close to 0 as possible. The output probability started as nearly even distribution and the destination rank probability gradually approaches to 1 while other approaches to 0. The feature learning balance the higher combination sorting and the non-repeated rule. So the network study the state data feature to reach the best sort for data in different word lines. 

\textbf{Test processing.}
The test for model is executed on some rand state data and then we statistic the score distribution with other baselines and rand sorting. Finally, the directly purpose of the network is to make the evaluation function value maximum, so we calculate the evaluation score according to equation \ref{equation:score} of data arranged with LSTM-resorting network,baselines method and original rand data. 

\section{EXPERIMENTS AND EVALUATION}
\subsection{Experiment statement}

We train and test our PDA-LSTM network with the RTX 4090 and relative parameters are as following in Table \ref{tab:exsetup}.
\begin{table}[]
    \centering
    \caption{Software platform configuration}
    \begin{tabular}{c|c}       
    \hline
    \textbf{Configuration}&\textbf{Setup information}\\
    \hline
        OS & Ubuntu 22.04 \\
        Frame & Pytorch 1.11.0 + CUDA 11.3\\
        VRAM&24GB\\
        GPU&GeForce RTX 4090\\
        CPU&Xeon(R) Platinum 8352V\\
        Memory&2GB \\ 
    \hline
    \end{tabular}
    \label{tab:exsetup}
\end{table}
We first do an experiment to identify the effectiveness of scoring function. It should sample a few data and calculation according to equation (\ref{equation:score}). Then test their BER at our experiment environment and make sure there is relationship between the BER and the LCM evaluation score. 
\par
In bit error rate (BER) testing experiment, we use the QLC NAND flash chips manufactured by Yangtze memory technologies with the model number of YMN0AQF1B1DCAD. The page size of the chip is 16KB data space + 2432B spare space. Logical page number of each block is 5544 and the block number of each plane is 403. The number of planes of each die is 4. The chip has 232 stack layers and each block has 6 strings. BER testing of the QLC 3D NAND flash chips is performed using the NanoCycler, a NAND flash characterisation instrument manufactured by NplusT. In the experiment, we used a random number pattern for program/erase cycles and the data retention temperature is set to 85\textdegree C. 
\par
Then we choose WBVM and DVDS with different code-length (CL) as the baselines to improve the advantage of arrangement for physical page data in LCM reduction. Fund on the physical page data adjusting, searching for the best way to obtain the best arrangement is also a signal task. So we change random sorting for data in row and line directions and based on the word line resorting, greedy and TSP model are also as comparison to the neural network. 
\par
Finally, funded on the conclusion about data arrangement impact on BER, we speculate that the reduction of bit error ratio can speed up the read and write operation. Flash read
latency is the total latency involved in flash sensing, data transferring, and LDPC decoding. So with the reduction of BER, ECC can spend less time and make the process more rapidly. To determine the performance gains due to data arrangement, we evaluated the performance of the system under different workloads using SSDsim.

\subsection{Effectiveness and necessary analysis}
\begin{figure}[htbp]
    \centering      \includegraphics[width=1.0\linewidth]{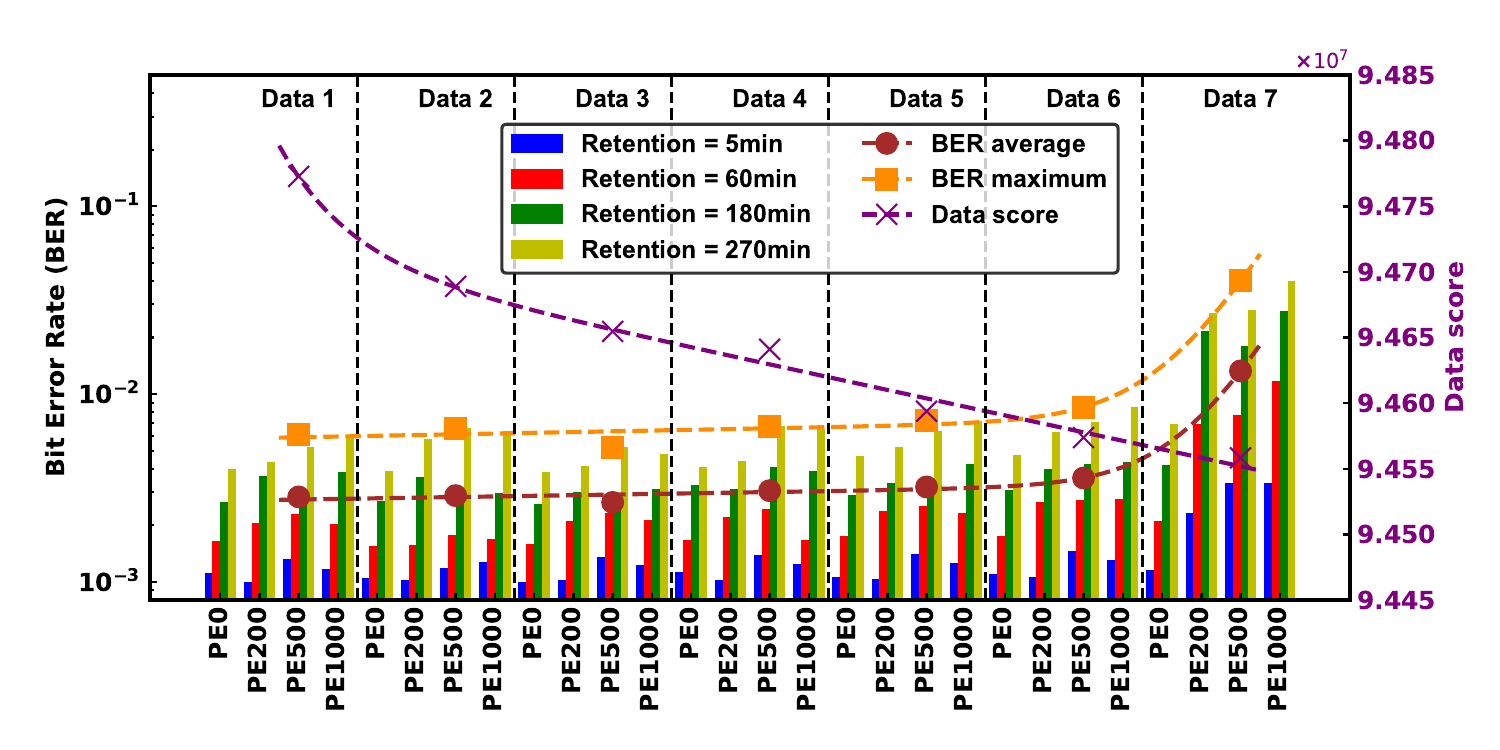}
    \caption{\textbf{Data score and BER relation.} Data pattern with higher score always performances lower BER in retention.}
    \label{fig:identify LCM}
\end{figure}

First, we analyzed the effectiveness of the scoring rules on the LCM effect and retention error characterization. Fig. \ref{fig:identify LCM} shows the data scores the data read BER in different retention time. It perfectly identify the scores depending on LCM evaluation function are inversely relative to the BER.
\par
\begin{figure*}
    \centering    \includegraphics[width=1\linewidth]{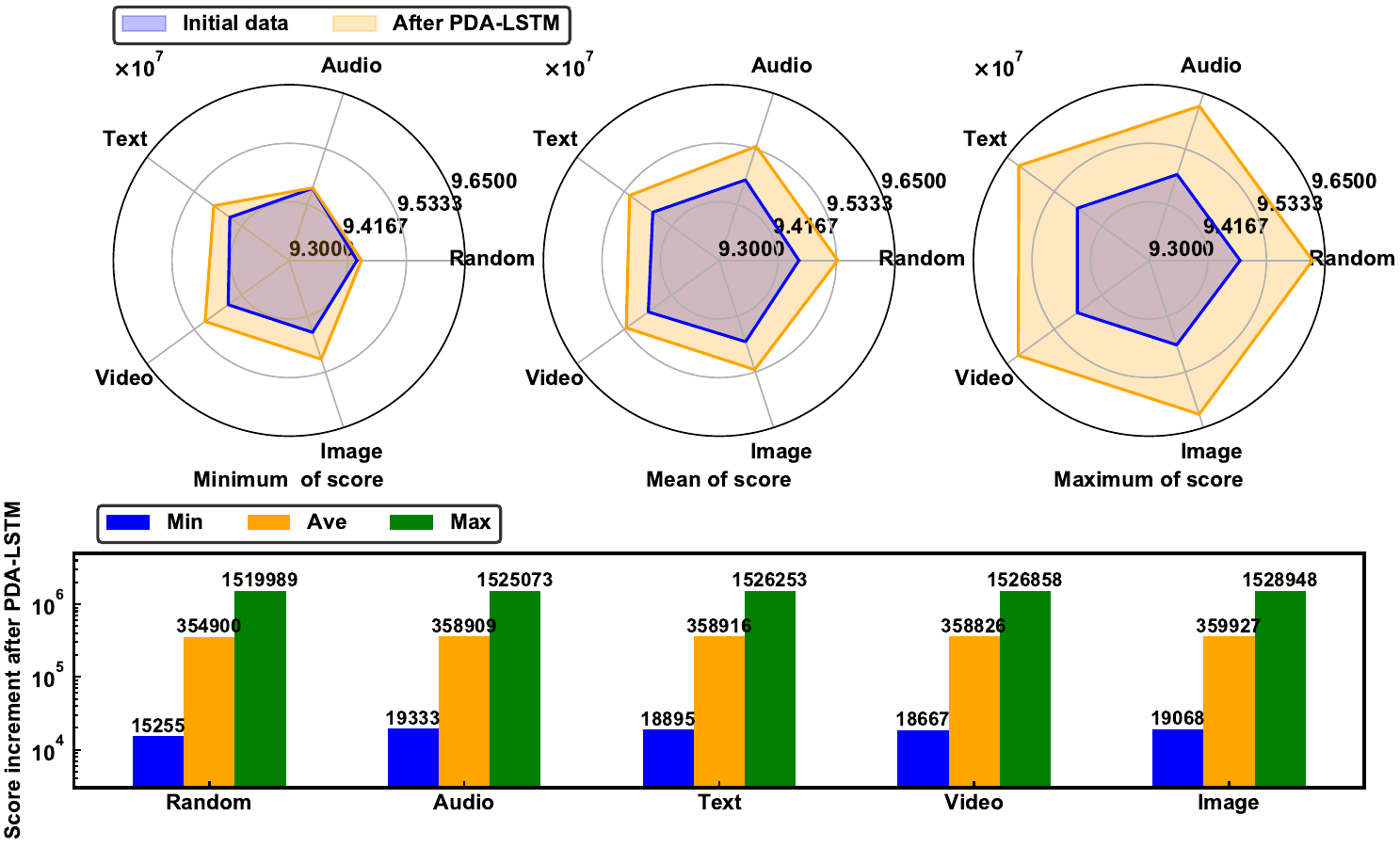}    \caption{Score increasement for different data types}
    \label{fig:rand_score}
\end{figure*}

Then, we analyzed the necessary for intelligent algorithm. To adjust the data state pattern, two adjusting modest can be applied. One moves the data to different bit lines to change the data pattern, another moves the data to different wordlines. To compare two adjusting effect, we try two kinds of movement randomly and statistic the score distribution, in Fig. \ref{fig:rand_score} (a), it is apparent that two methods results have little distance and movement to other word lines is more convenient\cite{mielke2017reliability}. So we finally select the data adjusting based on wordline positions because it is easier to execute in the same optimal results and can be easily implemented in SSD systems through address mapping.
\par
Alternatively, the exhaustive adjusting method is too tedious to find the optimal arrangement among all possible combinations. As is shown in Fig. \ref{fig:rand_score} (b), the simple random arrangement needs a lot of iterations to reach a higher score and there is still a big gap between the score and the best score. Moreover, score calculation remains a time-consuming operation. As a result, designing a method based on intelligent algorithms for a faster data arrangement is necessary. 

In intelligence methods, TSP should calculate the arranged data LCM impact score each iteration,which spend much time. And greedy method also should calculate the score and compare to find the best neighbour. So a neural network for data arrangement is the best choice because the calculation cost in its speculate process is much less than other traditional models.

\subsection{Model training and inference} 
\begin{figure*}
    \centering     \includegraphics[width=1\linewidth]{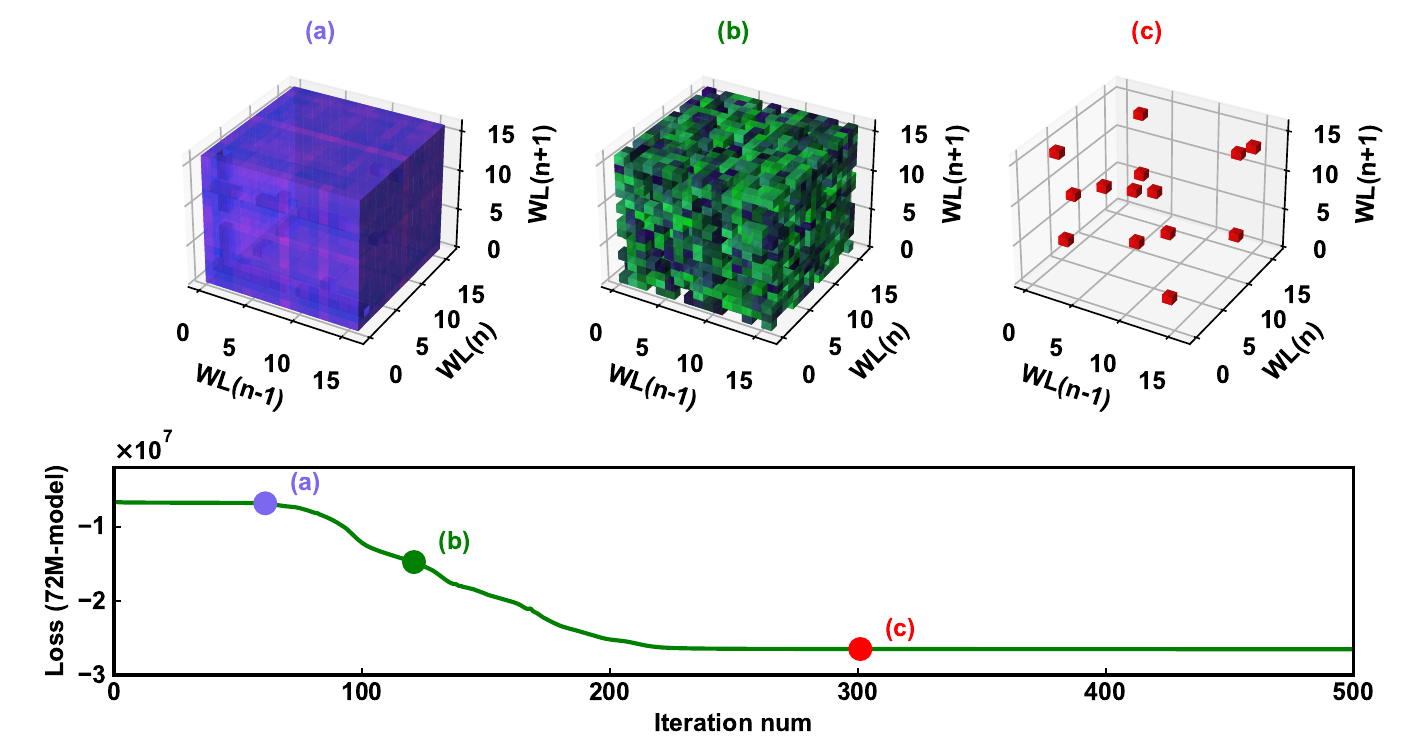}
    \caption{\textbf{Probability heat-map variation in training process.} From the left to the right represents the distribution probability heat-map of three adjacent page numbers respectively at the starting point of training, training 300 epoch and the end of training.}

    \label{fig:train change pub}
\end{figure*}

\begin{figure}
    \centering    \includegraphics[width=1.0\linewidth]{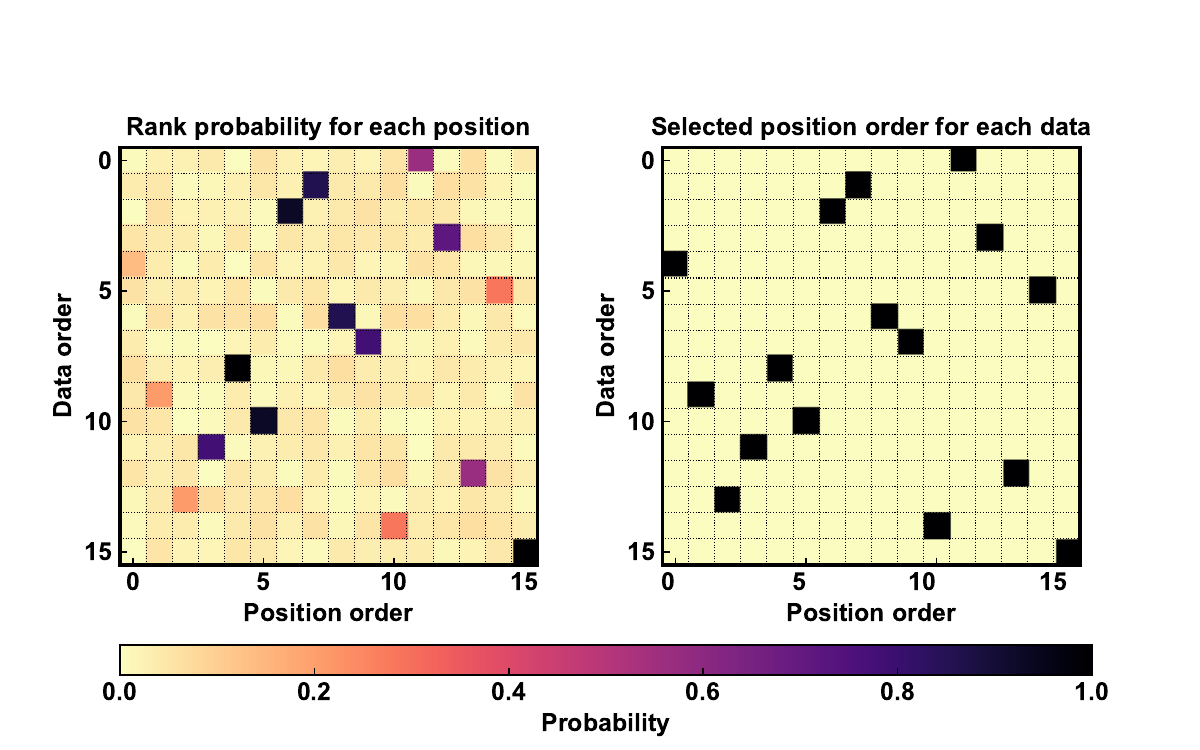}
    \caption{Visualization of output probability matrix}
    \label{fig:enter-label0}
\end{figure}

The output matrix of PDA-LSTM represents the propability of each physical in different wordlines to put in for each data. And through training process, it should assign the proper position larger and larger probability for appointed data. Fig. \ref{fig:train change pub} shows the output probability heat-maps respectively at the beginning, 500 epochs and the end of the training. It adequately indicates that in training iteration, the most appropriate triple page combinations probability are more and more closely to 1 while others are closely to 0.
The model inference each data to provide a path probability matrix, which is used to determine the data arrangement in the level of word lines. Commonly, we choose the largest probability index as the classification result, which is the darkest point in each row in this figure. Fig. \ref{fig:enter-label0} is one of the output probability matrix visual heat-map for a sample.

\subsection{Network Parameter size assignment based on calculation cost}
The model inference performance on hardware is the key to determine whether the model can be used effectively. For this model, we set four parameter sizes by adjust the network layers and hidden feature sizes of the LSTM layer. Then compare the data quality after arrangement from different models with various structure. 

\begin{figure}
    \centering
    \includegraphics[width=1.0\linewidth]{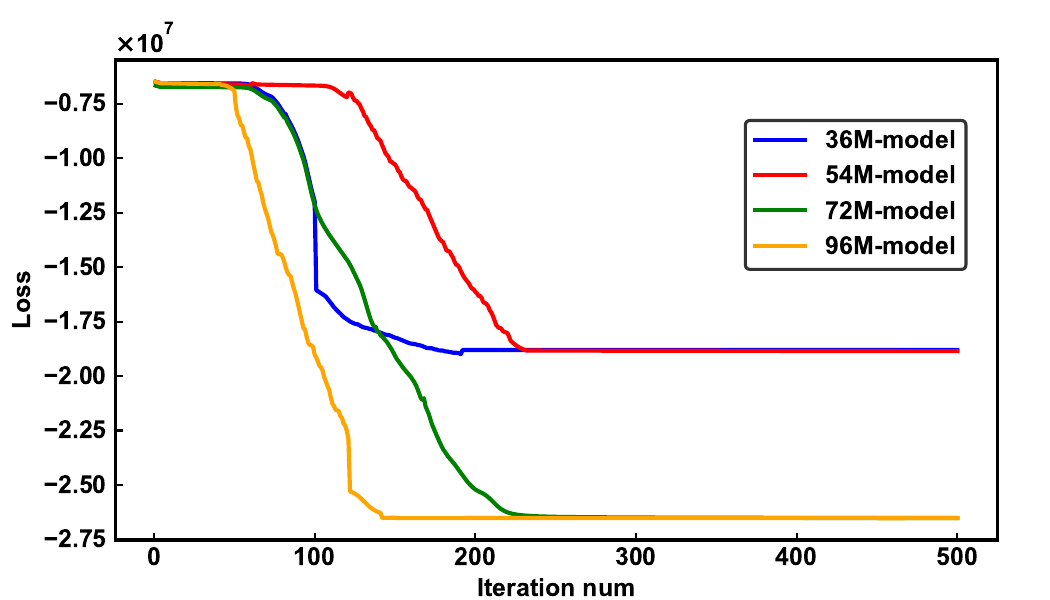}
    \caption{Loss curve for various structure network}
    \label{fig:parloss}
\end{figure}

We assert four kinds of structure for the network. Table \ref{tab:paraana} illustrates the hidden size, Linear layer quantity of four structure and their arranging data LCM evaluation scores. 
\begin{table}[]
    \centering
    \caption{Model structure feature and effect}
    \begin{tabular}{c|c|c|c}
        \hline
        Hidden size & Linear layer quantity&Parameter size &Score\\
        \hline
        16 & 1&36M&84677590 \\
        16 & 2&54M&84699956 \\
        32 & 1&72M&94953794 \\
        32 & 2&96M&94953798 \\ 
        \hline
    \end{tabular}
    \label{tab:paraana}
\end{table}
The effect of the model is increasing with the model parameter size until it reaches a critical value, when the size larger than the value, the arrangement it proposed is hardly more optimal. In addition, Fig. \ref{fig:parloss}, the learning curve reflects the loss changing during training for four different parameter sizes caused by structure variation, it shows that when selecting the model with parameter size of 72M, the training effect can reach the best and model with above 72M nearly will not propose a better result. 

In summary, we confirm to apply the model with hidden feature size of 32 in LSTM embedding layer and one Linear layer with 72M parameter size through experience and it can balance the calculation and the model effect.

\subsection{LCM evaluation score of arranged data from different methods}
\begin{figure}
    \centering    \includegraphics[width=1.0\linewidth]{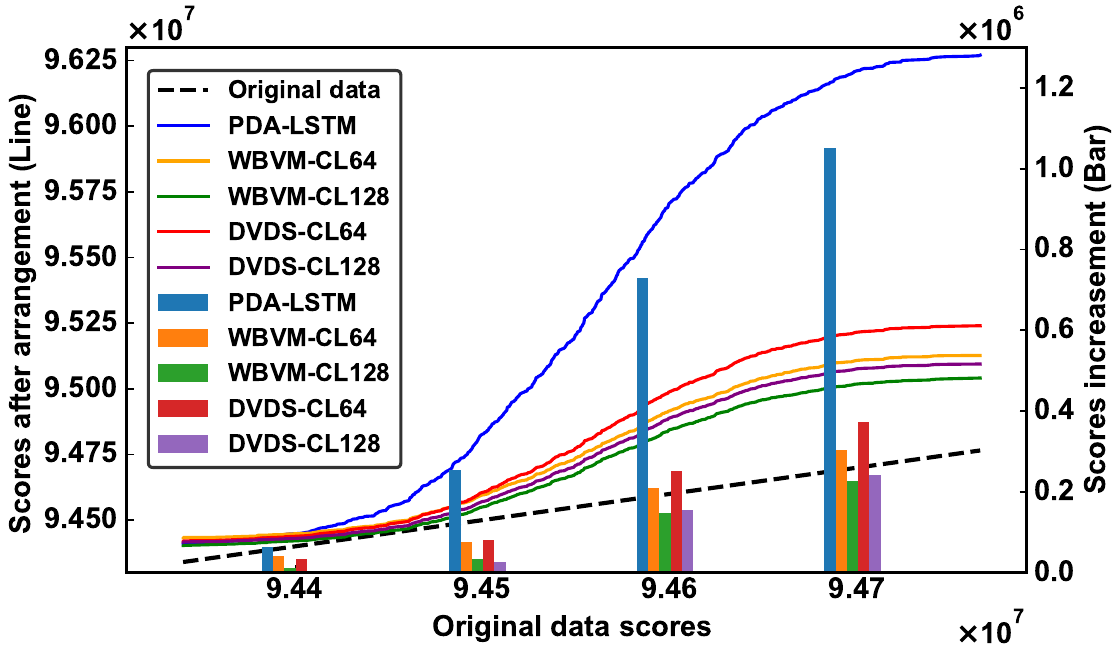}
    \caption{Data pattern scores with different processing methods}
    \label{fig:LCM score compare}
\end{figure}

In Fig. \ref{fig:LCM score compare}, we apply WBVM and DVDS with different code length (CL) as the baseline compared the LCM evaluation score with the PDA method. Code length is a hyperparameter of WBVM/DVDS, meaning the length of the dealing cell number. All method are compared with the data without any processing (Original data). With the increasement of the original data evaluation score, the PDA-LSTM method proposed more and more advantaged data arrangement result than others. 
\subsection{Arranged data BER from different methods}
\begin{figure}
    \centering      \includegraphics[width=1.0\linewidth]{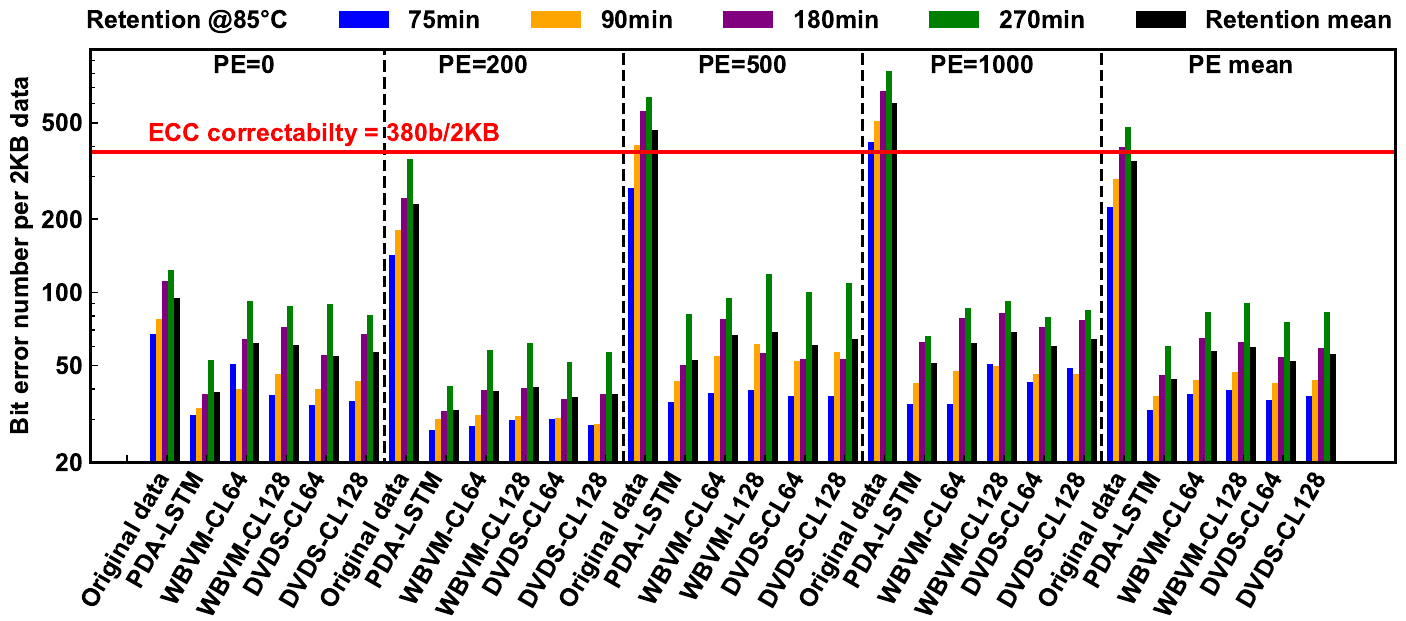}
    \caption{BER variation compare with PDA-LSTM}
    \label{fig:conclusion0}
\end{figure}

Fig. \ref{fig:conclusion0} is BER for different PEs during the retention time. With the PE cycles increasing, the distance between rand arrangement and other methods becomes more and more larger. With the retention time goes, the BER of data arranged with the prososed method is more and more less than that in others. So, we can find that our method for data arrangement can optimize the data retention character and improve the system reliability by reducing the BER. 

Based on Fig. \ref{fig:conclusion0}, we calculate the percentage optimization of the PDA-LSTM compared with other methods. We find that, PDA-LSTM reduces BER dramatically in different retention time, and the progress even can reach $40\%$ to $80\%$. PDA-LSTM reduces the average BER by 80.4\% compared with strategy without data arrangement, and by 21.4\%, 18.4\% compared with WBVM with code-length 64 and 128 respectively, and by 15.2\%, 20.4\% compared with DVDS with code-length 64 and 128 respectively. As a result, PDA-LSTM can notably decrease the read error for QLC 3D NAND flash. 
\par
\begin{figure}
    \centering       \includegraphics[width=1.0\linewidth]{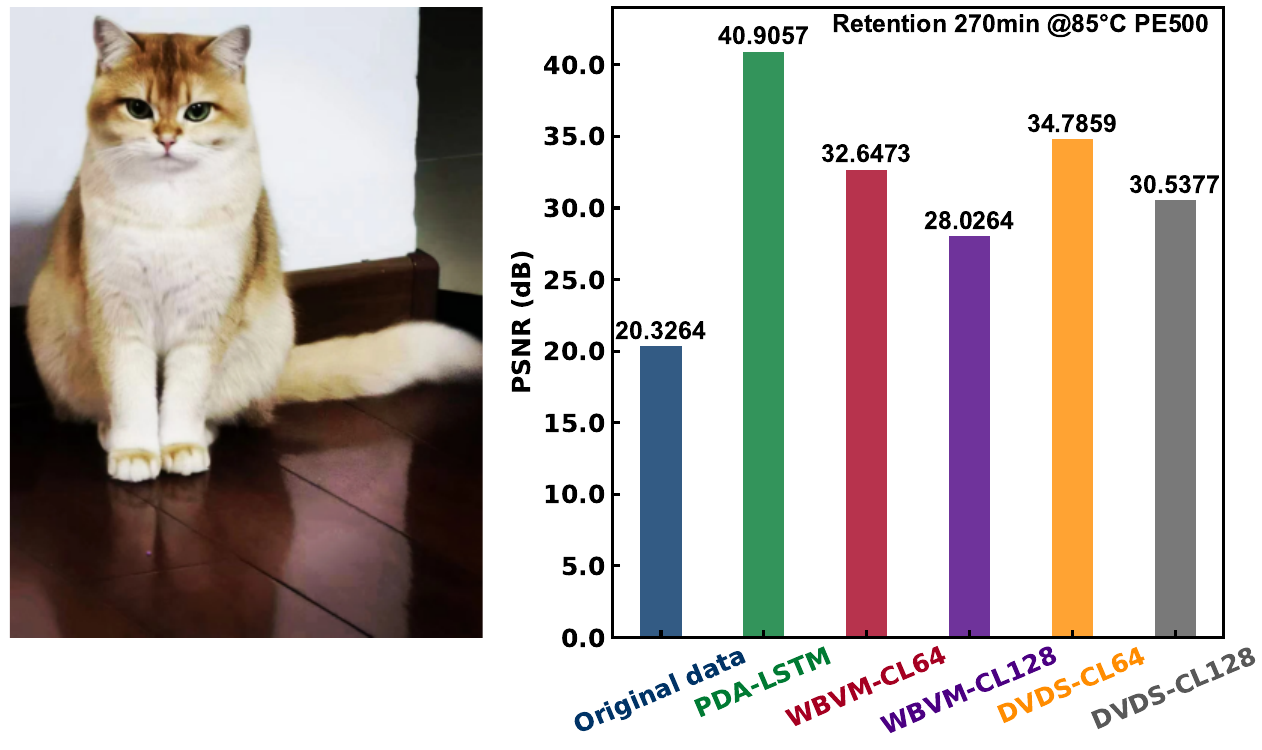}
    \caption{PSNR of image on QLC 3D NAND flash}
    \label{fig:mathvisual}
\end{figure}

We experiment on reality data to visualizate the BERs among different solutions. As shown in Fig. \ref{fig:mathvisual}, the left is the original image while the right is the Peak signal-to-noise ratio (PSNR) of the recovered image according to read data with different processing methods from QLC 3D NAND flash after 1000 PEs and 180 minutes retention time with 85\textdegree C. 

\begin{figure}
    \centering       \includegraphics[width=1.0\linewidth]{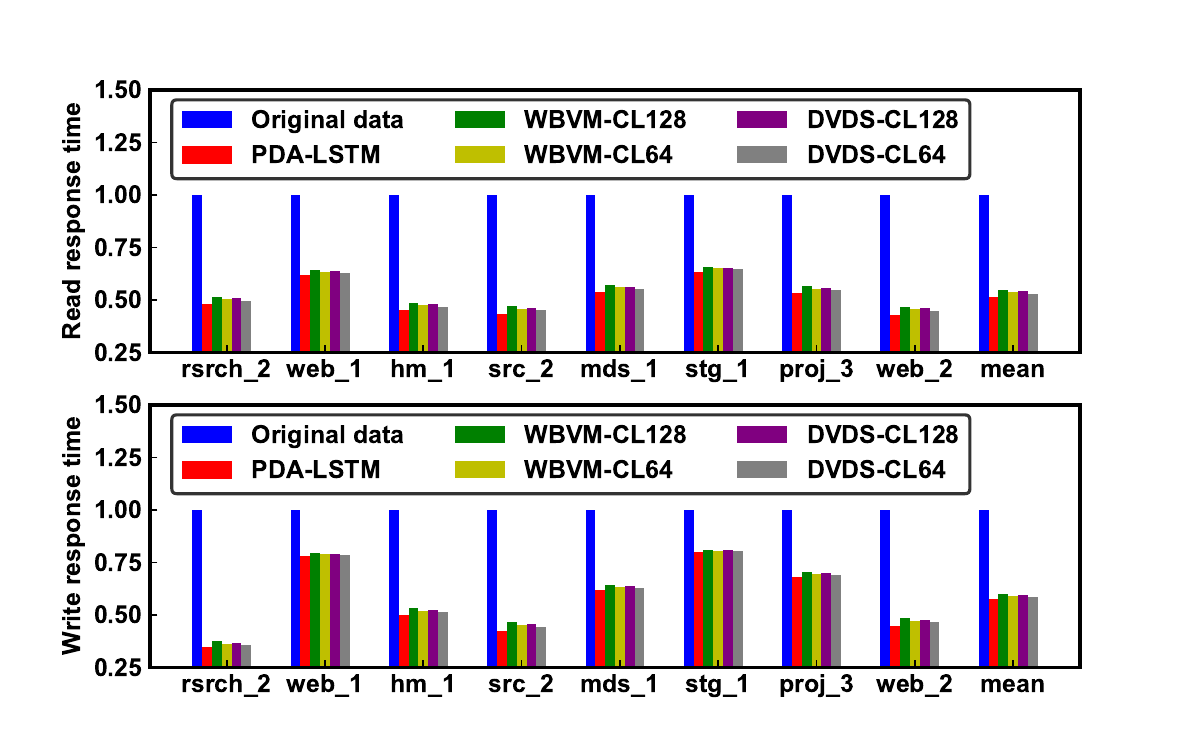}
    \caption{Read and write response time}
    \label{fig:SSDsim_result}
    
\end{figure}

\subsection{System performance analysis}
The BER reduction brings the decrease of execution operation time cost. The read and write operations of SSD are related to the time cost of ECC for error correction. Therefore, the decrease in BER can reduce the time required for ECC and improve read and write performance. We simulate eight benchmarks with SSD-Sim to compare the speed for our method and baselines.
SSD-Sim is a simulation tool based on modeling and simulation. It can simulate the storage mechanism and I/O operations of SSDs and various SSD configurations and workloads accurately by simulating SSD's storage mechanism and I/O operations. We compared the read and write response time of different LCM suppression methods under different benchmarks. 
It can be shown that PDA-LSTM make the process faster than other baselines. The advantages demonstrate mainly in write and read operations. The mean responded time of reading for PDA-LSTM data is only about 80$\%$ of data arranged with DVDS and 75$\%$ of that with WBVM when code length is 64.

\subsection{Comprehensive discuss}
Above all, it can fully describe the advantage of arranging the page data in decreasing the negative impact from LCM. The arranged data with PDA-LSTM obtains much lower BER than other strategies. Additionally, the method have a great advance when it executes on hardware platform because it needs less extra space to record the changes of data rearrangement than the other methods proposed formerly. Such as WBVM and DVDS, it needs a nearly 1/(code length) space to record the data transport when ours only need the double of word line number bytes. 
\begin{table}[]
    \centering
    \caption{The extra storage space for each method}
    \begin{tabular}{c|c|c}
       \hline
       Model & Redundant occupation & Score \\
       \hline
       Proposed PDA-LSTM  & 0 & 94953794 \\
       WBVM CL 64  &  BL$_{n}$/64 & 92407217 \\
       WBVM CL 128 &  BL$_{n}$/128 & 91731134 \\
       \hline
    \end{tabular}
    \label{tab:spaceana}
\end{table}

\section{CONLCUSION}
We proposed a page data arrangement method based on LSTM to mitigate LCM effect to reduce read errors. The PDA-LSTM model is proposed to complete page data arrangement to minimize the impact of LCM and improve the system reliability. Especially in cold data writing because that kind of data rarely written and it may be read for a long retention time. After completing the network training, score calculation, which is a time-consuming operation, is not required in the inference process. 
In inference process, the calculation cost and redundant space for address mapping are necessary. We hope that can be more precise LCM evaluation function and simplified models for some special reality data for better application in future works. 

\bibliographystyle{IEEEtran}
\bibliography{refs}

\begin{thebibliography}{10}
\providecommand{\url}[1]{#1}
\csname url@samestyle\endcsname
\providecommand{\newblock}{\relax}
\providecommand{\bibinfo}[2]{#2}
\providecommand{\BIBentrySTDinterwordspacing}{\spaceskip=0pt\relax}
\providecommand{\BIBentryALTinterwordstretchfactor}{4}
\providecommand{\BIBentryALTinterwordspacing}{\spaceskip=\fontdimen2\font plus
\BIBentryALTinterwordstretchfactor\fontdimen3\font minus \fontdimen4\font\relax}
\providecommand{\BIBforeignlanguage}[2]{{%
\expandafter\ifx\csname l@#1\endcsname\relax
\typeout{** WARNING: IEEEtran.bst: No hyphenation pattern has been}%
\typeout{** loaded for the language `#1'. Using the pattern for}%
\typeout{** the default language instead.}%
\else
\language=\csname l@#1\endcsname
\fi
#2}}
\providecommand{\BIBdecl}{\relax}
\BIBdecl

\bibitem{kim2021cmos}
M.-K. Kim, I.-J. Kim, and J.-S. Lee, ``{CMOS-compatible ferroelectric NAND flash memory for high-density, low-power, and high-speed three-dimensional memory},'' \emph{Science advances}, vol.~7, no.~3, p. eabe1341, 2021.

\bibitem{8013174}
Y.~Cai, S.~Ghose, E.~F. Haratsch, Y.~Luo, and O.~Mutlu, ``Error characterization, mitigation, and recovery in flash-memory-based solid-state drives,'' \emph{Proceedings of the IEEE}, vol. 105, no.~9, pp. 1666--1704, 2017.

\bibitem{9006406}
S.~Liang, Z.~Qiao, S.~Tang, J.~Hochstetler, S.~Fu, W.~Shi, and H.-B. Chen, ``An empirical study of {Quad-Level Cell (QLC) NAND Flash SSDs} for big data applications,'' in \emph{2019 IEEE International Conference on Big Data (Big Data)}, 2019, pp. 3676--3685.

\bibitem{liu_qlc_2017}
S.~Liu and X.~Zou, ``{QLC} {NAND} study and enhanced gray coding methods for sixteen-level-based program algorithms,'' vol.~66, pp. 58--66.

\bibitem{8739689}
Y.~Takai, M.~Fukuchi, R.~Kinoshita, C.~Matsui, and K.~Takeuchi, ``Analysis on heterogeneous ssd configuration with quadruple-level cell {(QLC) NAND} flash memory,'' in \emph{2019 IEEE 11th International Memory Workshop (IMW)}, 2019, pp. 1--4.

\bibitem{9631819}
Y.~Lee, J.~Yoon, K.~Lim, B.~Choi, G.-H. Park, J.~W. Jeon, J.-H. Bae, D.~M. Kim, D.~H. Kim, E.~Kwon, and S.-J. Choi, ``{Vertical and lateral charge losses during short time retention in 3-D NAND flash memory},'' in \emph{ESSDERC 2021 - IEEE 51st European Solid-State Device Research Conference (ESSDERC)}, 2021, pp. 279--282.

\bibitem{maconi2012comprehensive}
A.~Maconi, A.~Arreghini, C.~M. Compagnoni, A.~Spinelli, J.~Van~Houdt, A.~L. Lacaita \emph{et~al.}, ``{Comprehensive investigation of the impact of lateral charge migration on retention performance of planar and 3D SONOS devices},'' \emph{Solid-State Electronics}, vol.~74, pp. 64--70, 2012.

\bibitem{wang2020lateral}
F.~Wang, R.~Cao, Y.~Kong, X.~Ma, X.~Zhan, Y.~Li, and J.~Chen, ``{Lateral charge migration induced abnormal read disturb in 3D charge-trapping NAND flash memory},'' \emph{Applied Physics Express}, vol.~13, no.~5, p. 054002, 2020.

\bibitem{deguchi2017word}
Y.~Deguchi and K.~Takeuchi, ``{Word-line batch Vth modulation of TLC NAND flash memories for both write-hot and cold data},'' in \emph{2017 IEEE Asian Solid-State Circuits Conference (A-SSCC)}.\hskip 1em plus 0.5em minus 0.4em\relax IEEE, 2017, pp. 161--164.

\bibitem{deguchi2018write}
Y.~Deguchi, S.~Suzuki, and K.~Takeuchi, ``{Write and Read Frequency-Based Word-Line Batch VTH Modulation for 2-D and 3-D-TLC NAND Flash Memories},'' \emph{IEEE Journal of Solid-State Circuits}, vol.~53, no.~10, pp. 2917--2926, 2018.

\bibitem{suzuki2018endurance}
S.~Suzuki, Y.~Deguchi, T.~Nakamura, and K.~Takeuchi, ``{Endurance-based Dynamic VTH Distribution Shaping of 3D-TLC NAND Flash Memories to Suppress Both Lateral Charge Migration and Vertical Charge De-trap and Increase Data-retention Time by 2.7x},'' in \emph{2018 48th European Solid-State Device Research Conference (ESSDERC)}.\hskip 1em plus 0.5em minus 0.4em\relax IEEE, 2018, pp. 150--153.

\bibitem{marx_big_2013}
V.~Marx, ``The big challenges of big data,'' vol. 498, no. 7453, pp. 255--260, publisher: Nature Publishing Group.

\bibitem{6547630}
X.~Wu, X.~Zhu, G.-Q. Wu, and W.~Ding, ``{Data mining with big data},'' \emph{IEEE Transactions on Knowledge and Data Engineering}, vol.~26, no.~1, pp. 97--107, 2014.

\bibitem{7123563}
A.~Al-Fuqaha, M.~Guizani, M.~Mohammadi, M.~Aledhari, and M.~Ayyash, ``Internet of things: A survey on enabling technologies, protocols, and applications,'' \emph{IEEE Communications Surveys \& Tutorials}, vol.~17, no.~4, pp. 2347--2376, 2015.

\bibitem{ouyang2020optimization}
Y.~Ouyang, Z.~Xia, T.~Yang, D.~Shi, W.~Zhou, and Z.~Huo, ``{Optimization of performance and reliability in 3D NAND flash memory},'' \emph{IEEE Electron Device Letters}, vol.~41, no.~6, pp. 840--843, 2020.

\bibitem{khakifirooz202130}
A.~Khakifirooz, S.~Balasubrahmanyam, R.~Fastow, K.~H. Gaewsky, C.~W. Ha, R.~Haque, O.~W. Jungroth, S.~Law, A.~S. Madraswala, B.~Ngo \emph{et~al.}, ``{30.2 a 1tb 4b/cell 144-tier floating-gate 3d-nand flash memory with 40mb/s program throughput and 13.8 gb/mm 2 bit density},'' in \emph{2021 IEEE International Solid-State Circuits Conference (ISSCC)}, vol.~64.\hskip 1em plus 0.5em minus 0.4em\relax IEEE, 2021, pp. 424--426.

\bibitem{peng2020impacts}
X.~Peng, F.~Wang, Y.~Kong, M.~Jia, X.~Zhan, Y.~Li, and J.~Chen, ``{Impacts of lateral charge migration on data retention and read disturb in 3D charge-trap NAND flash memory},'' in \emph{2020 IEEE 15th International Conference on Solid-State \& Integrated Circuit Technology (ICSICT)}.\hskip 1em plus 0.5em minus 0.4em\relax IEEE, 2020, pp. 1--3.

\bibitem{mizoguchi2017lateral}
K.~Mizoguchi, S.~Kotaki, Y.~Deguchi, and K.~Takeuchi, ``{Lateral charge migration suppression of 3D-NAND flash by vth nearing for near data computing},'' in \emph{2017 IEEE International Electron Devices Meeting (IEDM)}.\hskip 1em plus 0.5em minus 0.4em\relax IEEE, 2017, pp. 19--2.

\bibitem{6044201}
A.~Maconi, A.~Arreghini, C.~M. Compagnoni, G.~Van~den bosch, A.~S. Spinelli, J.~Van~Houdt, and A.~L. Lacaita, ``{Impact of lateral charge migration on the retention performance of planar and 3D SONOS devices},'' in \emph{2011 Proceedings of the European Solid-State Device Research Conference (ESSDERC)}, 2011, pp. 195--198.

\bibitem{Wang_2020}
\BIBentryALTinterwordspacing
F.~Wang, R.~Cao, Y.~Kong, X.~Ma, X.~Zhan, Y.~Li, and J.~Chen, ``{Lateral charge migration induced abnormal read disturb in 3D charge-trapping NAND flash memory},'' \emph{Applied Physics Express}, vol.~13, no.~5, p. 054002, apr 2020. [Online]. Available: \url{https://dx.doi.org/10.35848/1882-0786/ab8729}
\BIBentrySTDinterwordspacing

\bibitem{8720566}
A.~Padovani, M.~Pesic, M.~A. Kumar, P.~Blomme, A.~Subirats, S.~Vadakupudhupalayam, Z.~Baten, and L.~Larcher, ``{Understanding and Variability of Lateral Charge Migration in 3D CT-NAND Flash with and Without Band-Gap Engineered Barriers},'' in \emph{2019 IEEE International Reliability Physics Symposium (IRPS)}, 2019, pp. 1--8.

\bibitem{8353632}
Y.~H. Liu, H.~Y. Lin, C.~M. Jiang, T.~Wang, W.~J. Tsai, T.~C. Lu, K.~C. Chen, and C.-Y. Lu, ``{Investigation of data pattern effects on nitride charge lateral migration in a charge trap flash memory by using a random telegraph signal method},'' in \emph{2018 IEEE International Reliability Physics Symposium (IRPS)}, 2018, pp. 6D.1--1--6D.1--5.

\bibitem{junger1995traveling}
M.~J{\"u}nger, G.~Reinelt, and G.~Rinaldi, ``The traveling salesman problem,'' \emph{Handbooks in operations research and management science}, vol.~7, pp. 225--330, 1995.

\bibitem{malek1989serial}
M.~Malek, M.~Guruswamy, M.~Pandya, and H.~Owens, ``Serial and parallel simulated annealing and tabu search algorithms for the traveling salesman problem,'' \emph{Annals of Operations Research}, vol.~21, pp. 59--84, 1989.

\bibitem{karabulut2014variable}
K.~Karabulut and M.~F. Tasgetiren, ``A variable iterated greedy algorithm for the traveling salesman problem with time windows,'' \emph{Information Sciences}, vol. 279, pp. 383--395, 2014.

\bibitem{graves2012long}
A.~Graves and A.~Graves, ``Long short-term memory,'' \emph{Supervised sequence labelling with recurrent neural networks}, pp. 37--45, 2012.

\bibitem{hochreiter1997long}
S.~Hochreiter and J.~Schmidhuber, ``Long short-term memory,'' \emph{Neural computation}, vol.~9, no.~8, pp. 1735--1780, 1997.

\bibitem{van2020review}
G.~Van~Houdt, C.~Mosquera, and G.~N{\'a}poles, ``A review on the long short-term memory model,'' \emph{Artificial Intelligence Review}, vol.~53, no.~8, pp. 5929--5955, 2020.

\bibitem{9241228}
M.~Zhang, F.~Wu, Q.~Yu, W.~Liu, Y.~Wang, and C.~Xie, ``{Exploiting Error Characteristic to Optimize Read Voltage for 3-D NAND Flash Memory},'' \emph{IEEE Transactions on Electron Devices}, vol.~67, no.~12, pp. 5490--5496, 2020.

\bibitem{8936476}
Q.~Li, L.~Shi, Y.~Cui, and C.~J. Xue, ``{Exploiting Asymmetric Errors for LDPC Decoding Optimization on 3D NAND Flash Memory},'' \emph{IEEE Transactions on Computers}, vol.~69, no.~4, pp. 475--488, 2020.

\bibitem{6392183}
N.~Y. Song and H.~Yan, ``{Autoregressive and Iterative Hidden Markov Models for Periodicity Detection and Solenoid Structure Recognition in Protein Sequences},'' \emph{IEEE Journal of Biomedical and Health Informatics}, vol.~17, no.~2, pp. 436--441, 2013.

\bibitem{8310323}
S.~Lee, C.~Kim, M.~Kim, S.-m. Joe, J.~Jang, S.~Kim, K.~Lee, J.~Kim, J.~Park, H.-J. Lee, M.~Kim, S.~Lee, S.~Lee, J.~Bang, D.~Shin, H.~Jang, D.~Lee, N.~Kim, J.~Jo, J.~Park, S.~Park, Y.~Rho, Y.~Park, H.-j. Kim, C.~A. Lee, C.~Yu, Y.~Min, M.~Kim, K.~Kim, S.~Moon, H.~Kim, Y.~Choi, Y.~Ryu, J.~Choi, M.~Lee, J.~Kim, G.~S. Choo, J.-D. Lim, D.-S. Byeon, K.~Song, K.-T. Park, and K.-h. Kyung, ``{A 1Tb 4b/cell 64-stacked-WL 3D NAND flash memory with 12MB/s program throughput},'' in \emph{2018 IEEE International Solid-State Circuits Conference - (ISSCC)}, 2018, pp. 340--342.

\bibitem{8895844}
Y.-H. Liu, T.-C. Zhan, T.~Wang, W.-J. Tsai, T.-C. Lu, K.-C. Chen, and C.-Y. Lu, ``{Investigation of Electron and Hole Lateral Migration in Silicon Nitride and Data Pattern Effects on ${V}_{{t}}$ Retention Loss in a Multilevel Charge Trap Flash Memory},'' \emph{IEEE Transactions on Electron Devices}, vol.~66, no.~12, pp. 5155--5161, 2019.

\bibitem{pasuto2013lateral}
A.~Pasuto, M.~Soldati \emph{et~al.}, ``{Lateral spreading.}'' in \emph{Treatise on geomorphology}.\hskip 1em plus 0.5em minus 0.4em\relax Academic Press, 2013, vol.~7, pp. 239--248.

\bibitem{9871606}
A.~Y. Al-Maliki and K.~Iqbal, ``{Improved Classification Accuracy of Hand Movements Using Softmax Classifier and Kalman Filter},'' in \emph{2022 44th Annual International Conference of the IEEE Engineering in Medicine \& Biology Society (EMBC)}, 2022, pp. 3191--3194.

\bibitem{8765616}
C.~Zambelli, R.~Bertaggia, L.~Zuolo, R.~Micheloni, and P.~Olivo, ``Enabling computational storage through {FPGA} neural network accelerator for enterprise {SSD},'' \emph{IEEE Transactions on Circuits and Systems II: Express Briefs}, vol.~66, no.~10, pp. 1738--1742, 2019.

\bibitem{9758404}
J.~Fan and T.~Lian, ``{Youtube Data Analysis using Linear Regression and Neural Network},'' in \emph{2022 International Conference on Big Data, Information and Computer Network (BDICN)}, 2022, pp. 248--251.

\bibitem{mielke2017reliability}
N.~R. Mielke, R.~E. Frickey, I.~Kalastirsky, M.~Quan, D.~Ustinov, and V.~J. Vasudevan, ``{Reliability of solid-state drives based on NAND flash memory},'' \emph{Proceedings of the IEEE}, vol. 105, no.~9, pp. 1725--1750, 2017.

\end{thebibliography}


 




\vfill

\end{document}